\documentclass{elsart}

\usepackage{rotating,epsfig,graphicx,amssymb,amsmath}
\begin{document}
\runauthor{G. Litak at al.}

\begin{frontmatter}

\title{
Intermittent behaviour of a Cracked Rotor in the resonance region
}
\author[Lublin]{Grzegorz Litak
,}
\author[Cleveland]{Jerzy T. Sawicki\thanksref{E-mail2}}
\address[Lublin]{Department of Applied Mechanics, Technical University of
Lublin,  Nadbystrzycka~36, PL-20-618 Lublin, Poland}
\address[Cleveland]{
Cleveland State University, Department of Mechanical Engineering, Cleveland,~OH~44115
}

\thanks[E-mail2]{E-mail:
j.sawicki@csuohio.edu (J.T. Sawicki)}

\begin{abstract}
Vibrations of the Jeffcott rotor are modelled by a three degree of freedom system 
including 
coupling between  lateral and torsional modes.  The 
crack in a rotating shaft of
the rotor  is introduced  via time dependent stiffness with off diagonal 
couplings.
Applying the external torque to the system allows  to  observe the 
effect of crack "breathing" and gain insight into the system.  It is manifested
in the complex dynamic behaviour of the rotor in the region of  internal resonance, 
showing a quasi--periodic 
motion 
or even non-periodic  behaviour. In the present paper report,  we show the system 
response to the external torque excitation
using nonlinear analysis tools such as bifurcation diagram, phase portraits, 
Poincar\'e 
maps and wavelet
power spectrum. In the region of resonance we study intermittent motions 
based on laminar phases interrupted by a series nonlinear beats.
\end{abstract}


\end{frontmatter}

\section{Introduction}
In many technical devices and machines possessing rotors, i.e.  
turbines, pumps, compressors, etc., their  
actual dynamic condition determines their proper and safe operation. Often, these 
machines have 
 to work over the extended periods of time 
in various temperature regimes under large variable 
loading. As a consequence, the components of the machine are exposed to potential 
damage, 
such as for example transverse crack \cite{gash1976,gash1993}.
For 
reliable and safe operation of such machines
their rotors should be systematically monitored for 
presence of
cracks and 
tested  using non--destructive tests or vibration based 
techniques \cite{sawicki2002}. Structural health assessment of the rotating 
components 
may be performed by examining the dynamical response of the system subjected 
to external excitation applied to the rotating shaft.
Such a  strategy  requires solution and  study of the dynamical behaviour of 
the rotor \cite{dimarogonas1996,wauer1990a}. The main difference between the 
cracked and uncracked 
rotors can
appear in the region of internal or combination resonances \cite{plaut1995,chondros1997}.
One should notice that the dynamics of the rotor is related to the 
coupling of 
torsional and lateral vibrations and the whole system is 
nonlinear. The crucial point is related to the model of the "breathing" crack, 
indicating that cracks involve additional parametric excitations with the 
frequency equal to an angular velocity of the rotor rotation 
\cite{penny2002,sawicki2005}.   
The crack introduces extra nonlinearity terms and 
produces coupling terms of lateral and torsional modes.

\begin{figure}[htb]
\center{
 \includegraphics[width=12.0cm,angle=0]{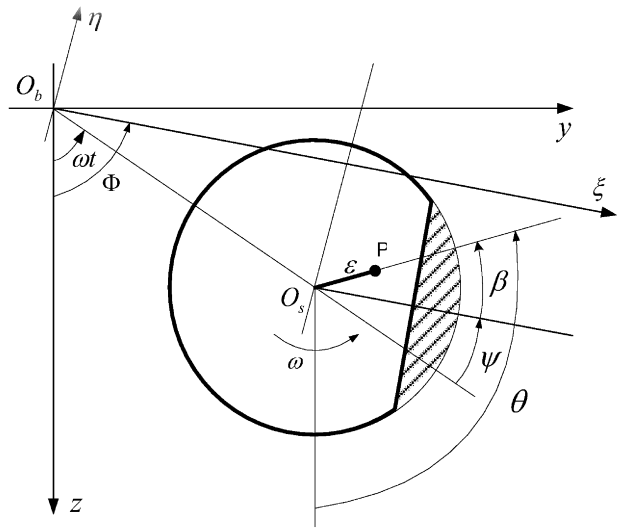}}

\caption{ \label{fig1} Schematic plot  of the cracked Jeffcott rotor \cite{sawicki2005}.
}
\end{figure}

Tondl \cite{tondl1965}, Cohen and
Porat \cite{cohen1985} and Bernasconi \cite{bernasconi1987} showed that typical
lateral excitations, such as
unbalance, may result in both lateral and
torsional responses of uncracked rotor.  Continuing these works Muszynska et al.
\cite{muszynska1992} and
Bently {\em et al.} \cite{bently1997} discussed rotor coupled
lateral and torsional vibrations due to unbalance.
Their experimental results exhibited the existence of significant torsional
vibrations, due to coupling with the lateral
modes.
Further  reported works
have used this coupling under external excitation as a means to identify the
presence of cracks \cite{collins1991,darpe2004}.
It has become  clear that the spectral components of the shaft response under
radial
excitation
can be used to identify
the presence of transverse shaft cracks. In particular, Iwatsubo {\em et al.}
\cite{iwatsubo1992}
considered the vibrations of a slowly
rotating shaft subject to either periodic or impulsive excitation. They identified
specific harmonics in the
response spectrum, which are combinations between the rotation speed and excitation
frequency and  they can be
used to detect the presence of the crack. Also, Iwatsubo {\em et al.}
\cite{iwatsubo1992} noted that the
sensitivity of the response to the
magnitude of the damage depends on the value of the excitation frequency chosen for
the detection. Finally,
Ishida and Inoue \cite{ishida2001} considered the response of a horizontal rotor to
harmonic external torques.
Furthermore, Sawicki et al. \cite{sawicki2003a,sawicki2003b,sawicki2005} examined
transient response of the
cracked rotor,
including the stalling effect under the
constant driving torque.

In this paper we continue these investigations. By applying the external harmonic
torque excitation we  show the characteristic features
of internal resonances leading to quasi-periodic and non--periodic types of
motion.

\section{The model and equations of motion}

The model of the Jeffcott rotor, which is by a definition limited to
a rotating mass disc and massless rotating elastic shaft, is presented in Fig. 1.

The angular position of the unbalance vector can be written as
\begin{eqnarray}
&& \theta (t ) = \omega t + \psi (t ) + \beta,
\label{eq1}
\end{eqnarray}
 where  $\omega$ is a
constant spin
speed of the shaft,
  $\psi(t )$ is a torsional angle, while $\beta$  is the fixed angle between the
unbalance
vector and
the
centerline
of the transverse shaft surface crack. It should be noted that
\begin{eqnarray}
\dot \theta =  \dot \Phi = \omega + \dot \psi, \hspace{1cm}
\ddot \theta = \ddot \psi,
\label{eq2}
\end{eqnarray}
where $\Phi$ is the spin angle of the rotor (Fig. 1).

The kinetic $T$ and potential  $U$ energies for the rotor system subjected
to lateral and torsional vibrations
can be
expressed as
follows:
\begin{eqnarray}
&& T= \frac{1}{2} J_p \dot \theta^2 +  \frac{1}{2} M \left( \dot z^2 + \dot y^2  \right) +
\frac{1}{2} M \epsilon^2 \dot \theta^2 + M\epsilon \dot \theta  \left( -\dot z \sin \theta + \dot y
\cos
\theta \right) \label{eq3} \\
&& U=  \frac{1}{2} \left\{ z~y \right\} {\bf K}_I \left\{ \begin{array}{c}
z \\ y
\end{array}\right\} + \frac{1}{2} k_t \psi^2, \label{eq4}
\end{eqnarray}
where the stiffness matrix  reads
\begin{eqnarray}
&& {\bf K}_I=\left[ \begin{array}{cc} k_{zz} & k_{zy} \\
 k_{yz} & k_{yy} \end{array}
\right].
\label{eq5}
\end{eqnarray}

The coupled equations for transverse and
torsional
motion for the rotor system take the following form:
\begin{eqnarray}
&& M \ddot z + C_l \dot z + k_{zz} z + k_{zy} y = F_z + M \epsilon \left( \dot \theta^2 \cos \theta + \ddot
 \theta \sin \theta \right) \label{eq6}\\
&& M \ddot y + C_t \dot z + k_{yz} z + k_{yy} y = F_y + M \epsilon \left( \dot \theta^2 \sin \theta + \ddot
 \theta \cos \theta \right) \label{eq7} \\
&& J_p \ddot \psi + C_t \dot \psi + \epsilon C_l ( \dot z \sin \theta
- \dot y \cos \theta ) -  \epsilon ( F_z \sin \theta - F_y \cos \theta) \label{eq8} \\
&&
+ \epsilon \left[ ( k_{zz} z +k_{zy} y) \sin \theta
- (k_{yz} z + k_{yy} y)
\cos \theta \right] +\frac{ \partial U}{\partial \psi} = T_e,  \nonumber
\end{eqnarray}
where $M$ and $J_p$ is the mass and mass moment of inertia of the disk, while
$\epsilon$ is eccentricity of the disk.
  $F_z$ and
$F_y$ are the external forces (including gravity) in $z$ and
$y$
directions, respectively, $C_l$ and  $C_t$ are lateral and  torsional damping
coefficients,
$T_e$ is an external torque.
In our case
\begin{eqnarray}
&& F_z=-Mg,~~~~ F_y=0,~~~~ T_e=T_0\sin(\omega_et).
\label{eq9}
\end{eqnarray}
where $g$ is the gravitational acceleration, $T_0$ and $\omega_e$ are amplitude and
frequency of
the external torque, respectively.

The stiffness matrix for a Jeffcott rotor with a cracked shaft in rotating
coordinates can be written as:
\begin{eqnarray}
&& {\bf K}_R=\left( \begin{array}{cc} k_{\xi} & 0 \\ 0 & k_{\eta}
\end{array} \right)= \left( \begin{array}{cc} k & 0 \\ 0 &
k \end{array} \right) - f(\Phi) \left( \begin{array}{cc} \Delta k_{\xi} & 0 \\ 0 &
\Delta k_{\eta}
\end{array} \right),
\label{eq10}
\end{eqnarray}

where the first matrix, on the left hand side, refers to the stiffness of the uncracked shaft, and
the
second defines the variations in shaft
stiffness $k_{\xi}$ and $k_{\eta}$ in $\xi$ and $\eta$  directions, respectively.
The function $f ( \Phi)$ is a
crack
steering function which
depends on the angular position of the crack $\Phi$.
The hinge model of the crack might be an appropriate representation for very small
cracks,
Mayes and Davies \cite{mayes1984} proposed a
model with a smooth transition between the opening and closing of the crack that is
more adequate for larger cracks. In
this case the crack steering function, or the Mayes function, takes the
following form:
\begin{eqnarray}
&& f( \Phi)=\frac{1+\cos (\Phi)}{2}.
\label{eq11}
\end{eqnarray}

The stiffness matrix for a Jeffcott rotor with a cracked shaft in inertial
coordinates, ${\bf K}_I$ is derived as

\begin{eqnarray}
{\bf K}_I={\bf T} {\bf K}_R {\bf T}^{-1}, ~~~ {\bf T}=\left( \begin{array}{cc}
\cos \Phi & - \sin \Phi \\
\sin \Phi & \cos \Phi
\end{array} \right),
\end{eqnarray}
where
\begin{eqnarray}
\Delta k_1 = \frac{ \Delta k_{\xi} + \Delta_{\eta}}{2}, ~~~~~ \Delta k_1 = \frac{
\Delta k_{\xi} - \Delta_{\eta}}{2}.
\end{eqnarray}

Consequently after \cite{sawicki2005}:
\begin{eqnarray}
&& \frac{\partial U}{\partial \psi}= - \frac{k}{4} \frac{\partial f(
\Phi)}{\partial \psi} \left[ (\Delta k_1 + \Delta k_2 \cos 2 \Phi) z^2
\right. \nonumber \\
&&  \left. +2zy \Delta
k_2 \sin 2 \Phi +(\Delta k_1 - \Delta k_2 \cos 2 \Phi) y^2   \right] \\ && +
\frac{f
(\Phi) k \Delta k_2}{2} \left[ z^2 \sin 2 \Phi -2 zy \cos 2 \Phi - y^2 \sin 2
\Phi \right]  + k_t \psi. \nonumber \end{eqnarray}

\section{Results of Simulations}

Assuming  the example of simply supported
steel shaft, with a disc at the center
and a  transverse crack  located near the disk,
 we applied  Eqs. \ref{eq6}-\ref{eq8} and performed the simulations.
The parameters of the examined system are shown in the
Table 1.
Simulations were done using the Euler integration procedure with a integration
step $\delta t=2\pi/(36000\omega)$ and a sampling time $\Delta t =1000 \delta t$.

\begin{figure}[htb]
\center{
 \includegraphics[width=9.5cm,angle=-90]{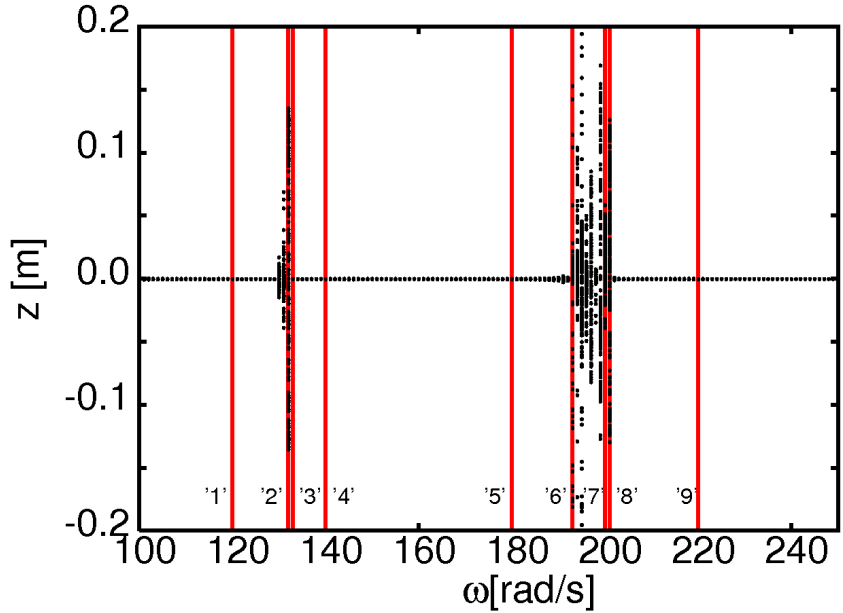}}

\caption{ \label{fig2} Bifurcation diagram of the cracked rotor.
The numbers '1'--'9'
correspond to $\omega=$  120, 132, 133, 140, 180, 193, 200, 201, and 220 rad/s.
}
\end{figure}

\begin{figure}[htb]

\includegraphics[width=5.6cm,angle=-90]{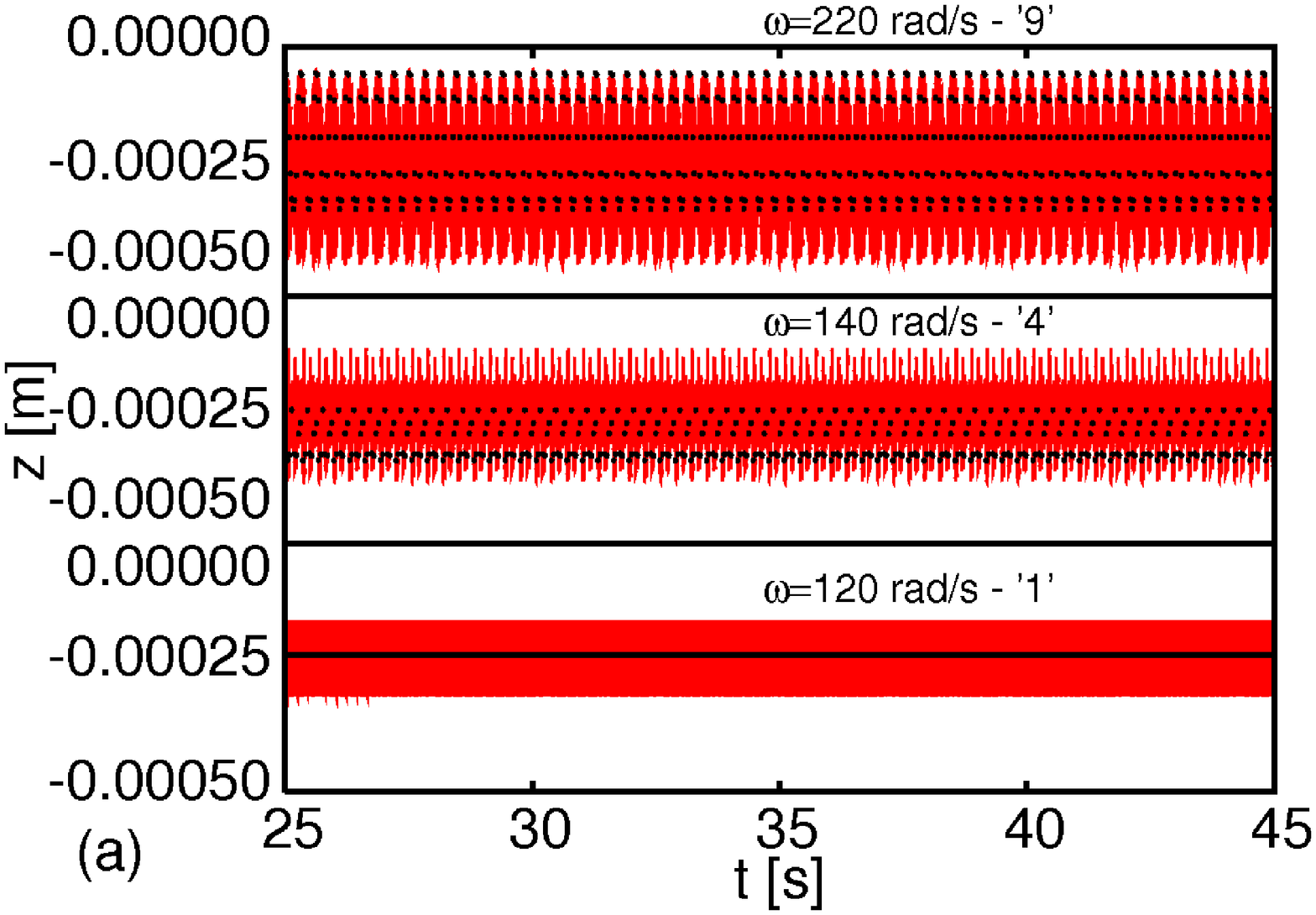}
\hspace{-1.5cm}
\includegraphics[width=5.6cm,angle=-90]{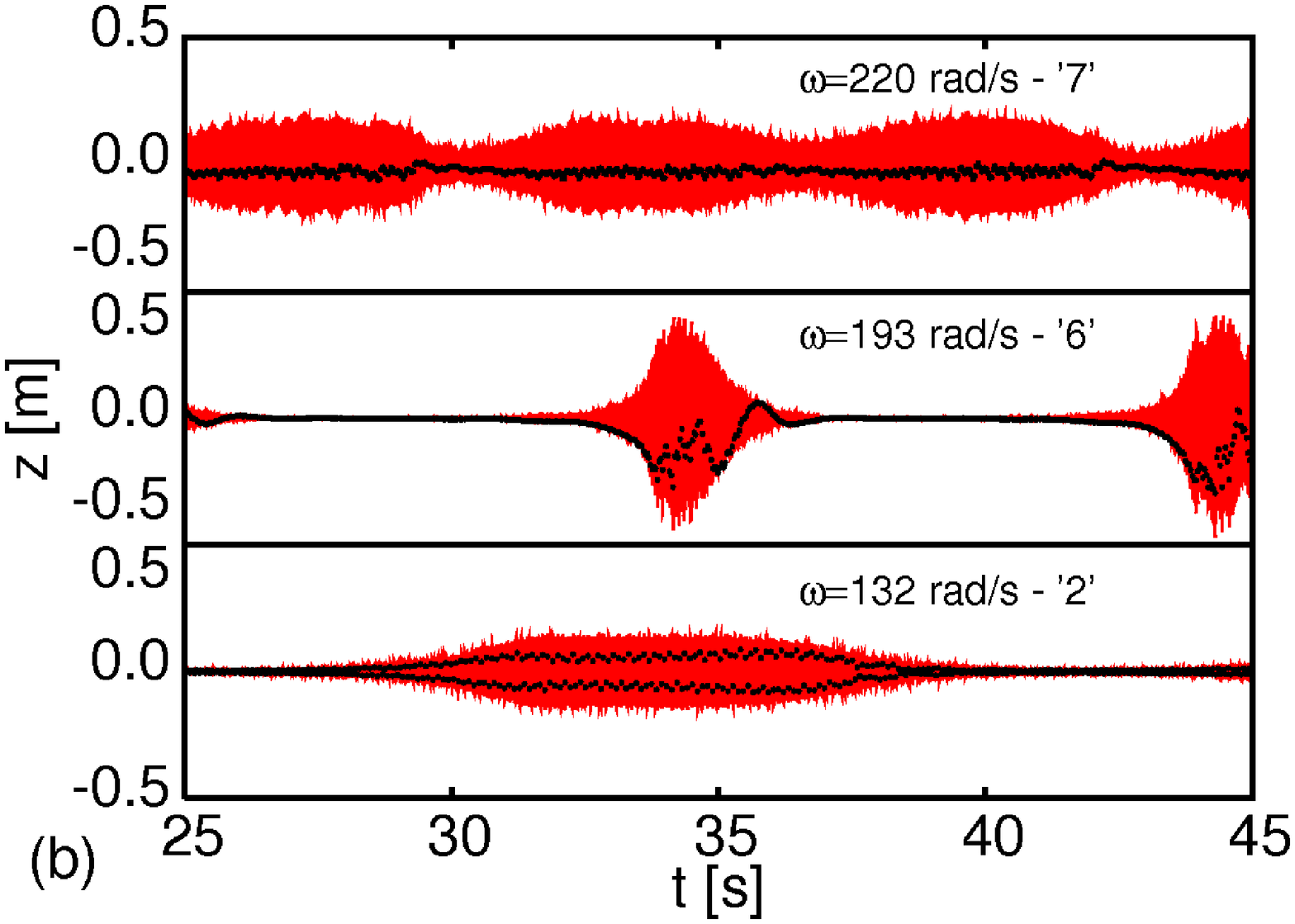}

\caption{ \label{fig3} Time histories of displacement in the vertical direction (a) for $\omega=120$, 140, 220
rad/s
(small 
amplitude
periodic vibrations)
and  (b)  for $\omega=132$, 200, 201 rad/s  (intermittent vibrations). Black points shows
stroboscopic
points. Numbers '1', '4', '9', '2', '6', and '7'  indicate the corresponding values in the bifurcation diagram
 (Fig. 2).
}
\end{figure}

\begin{figure}[htb]

\hspace{0.5cm} \includegraphics[width=5.1cm,angle=-90]{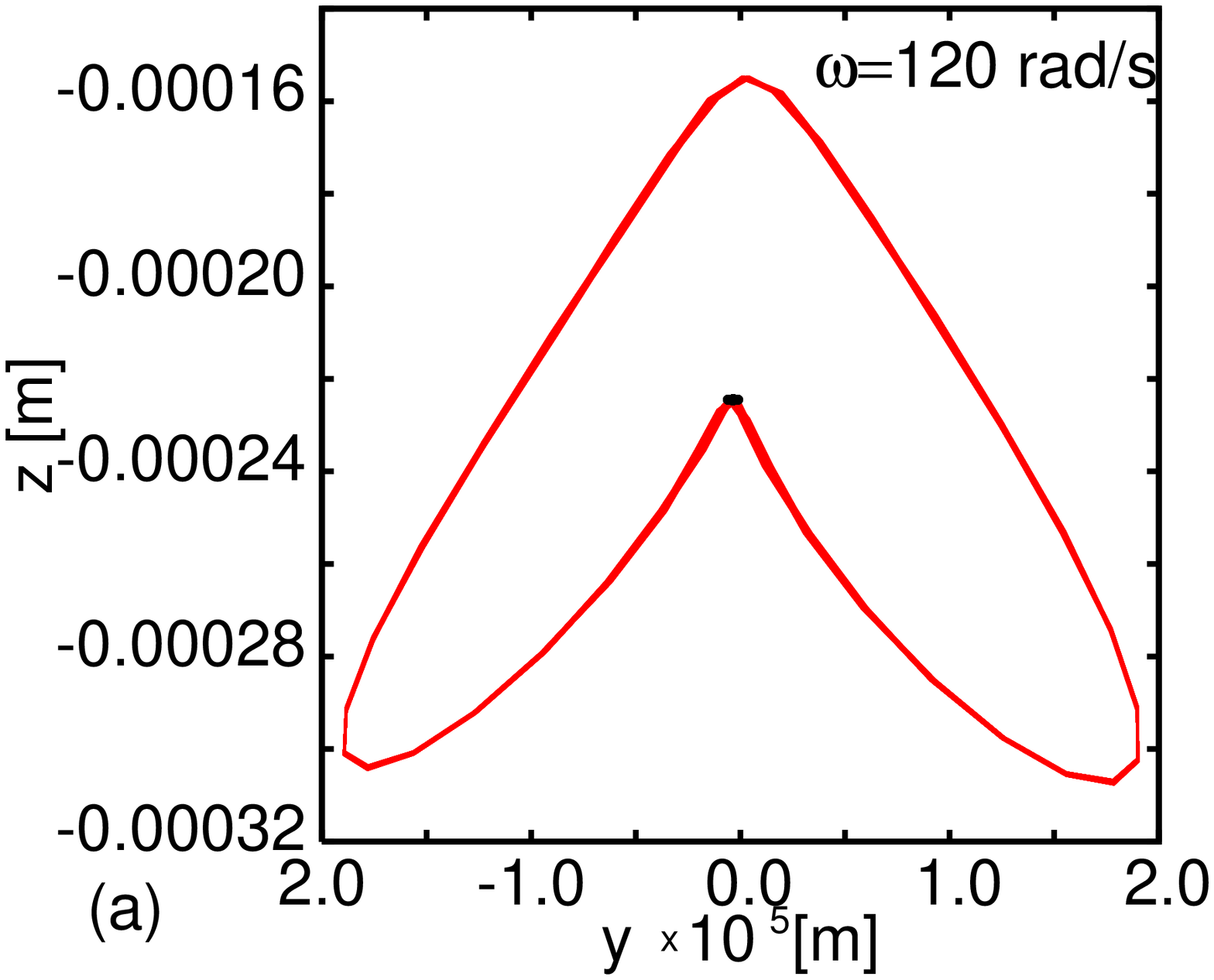}
\hspace{-0.5cm}
\includegraphics[width=5.1cm,angle=-90]{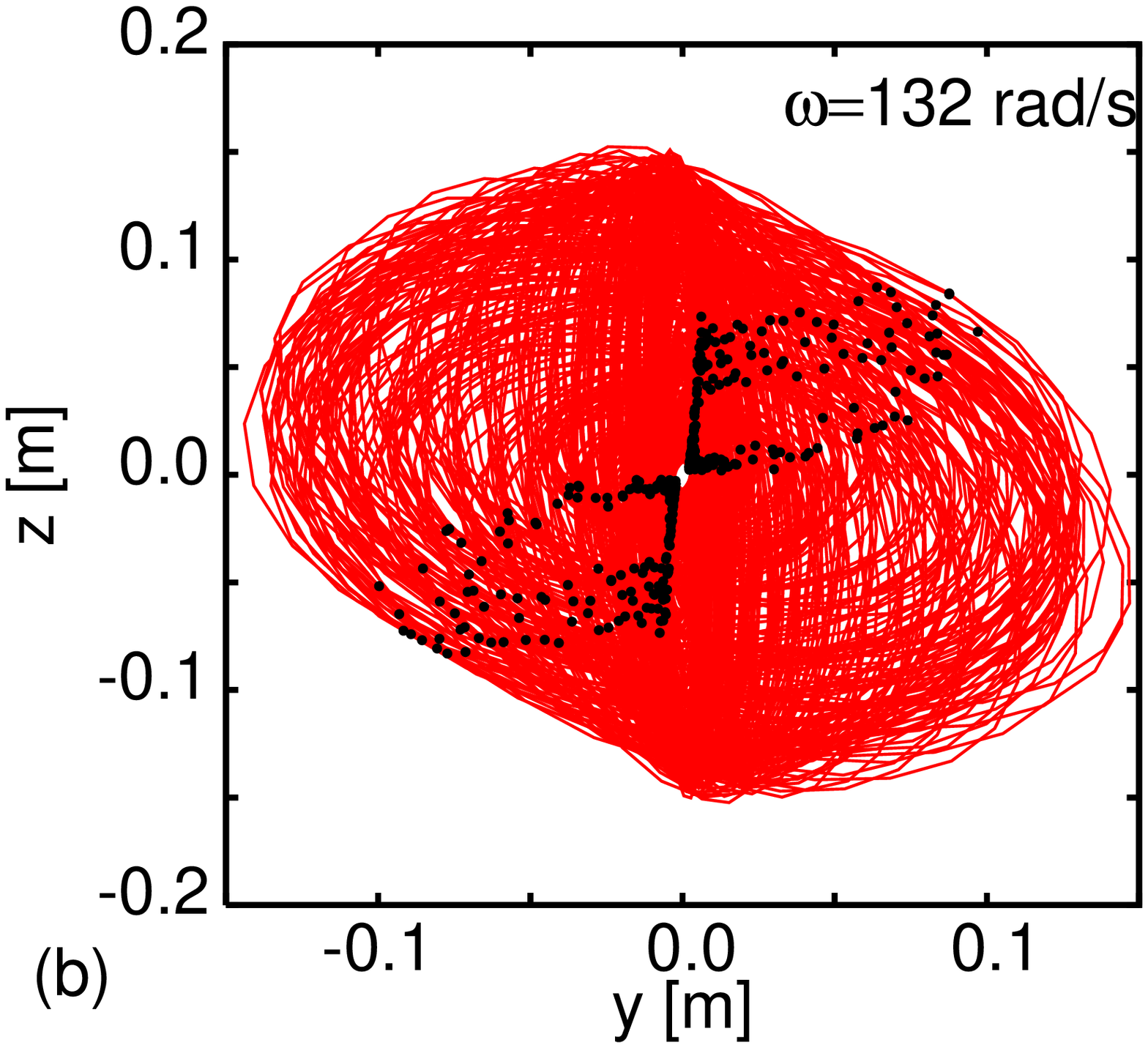}

\vspace{-0.3cm} 
\hspace{0.5cm}
\includegraphics[width=5.1cm,angle=-90]{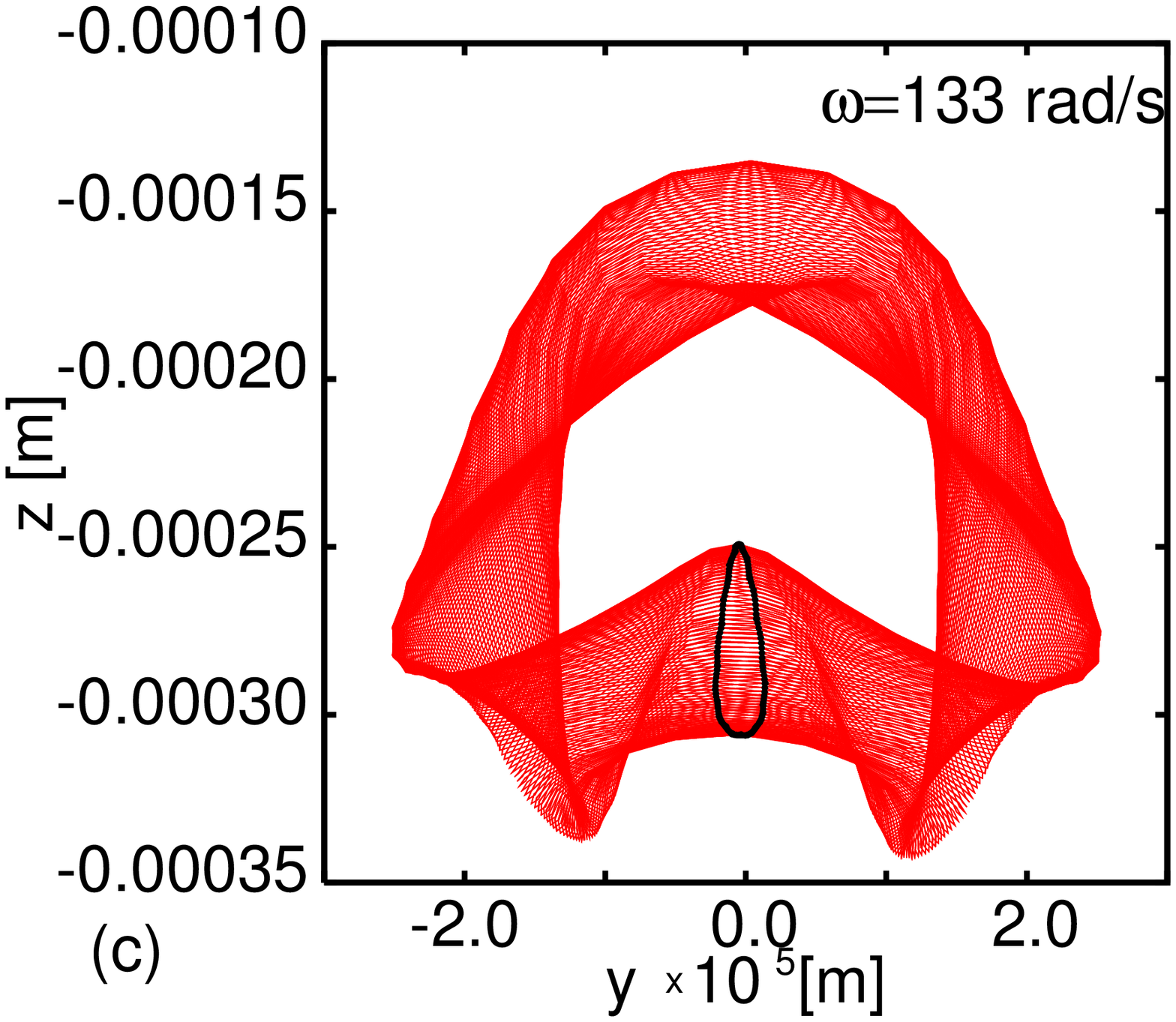} \hspace{-0.5cm}
\includegraphics[width=5.1cm,angle=-90]{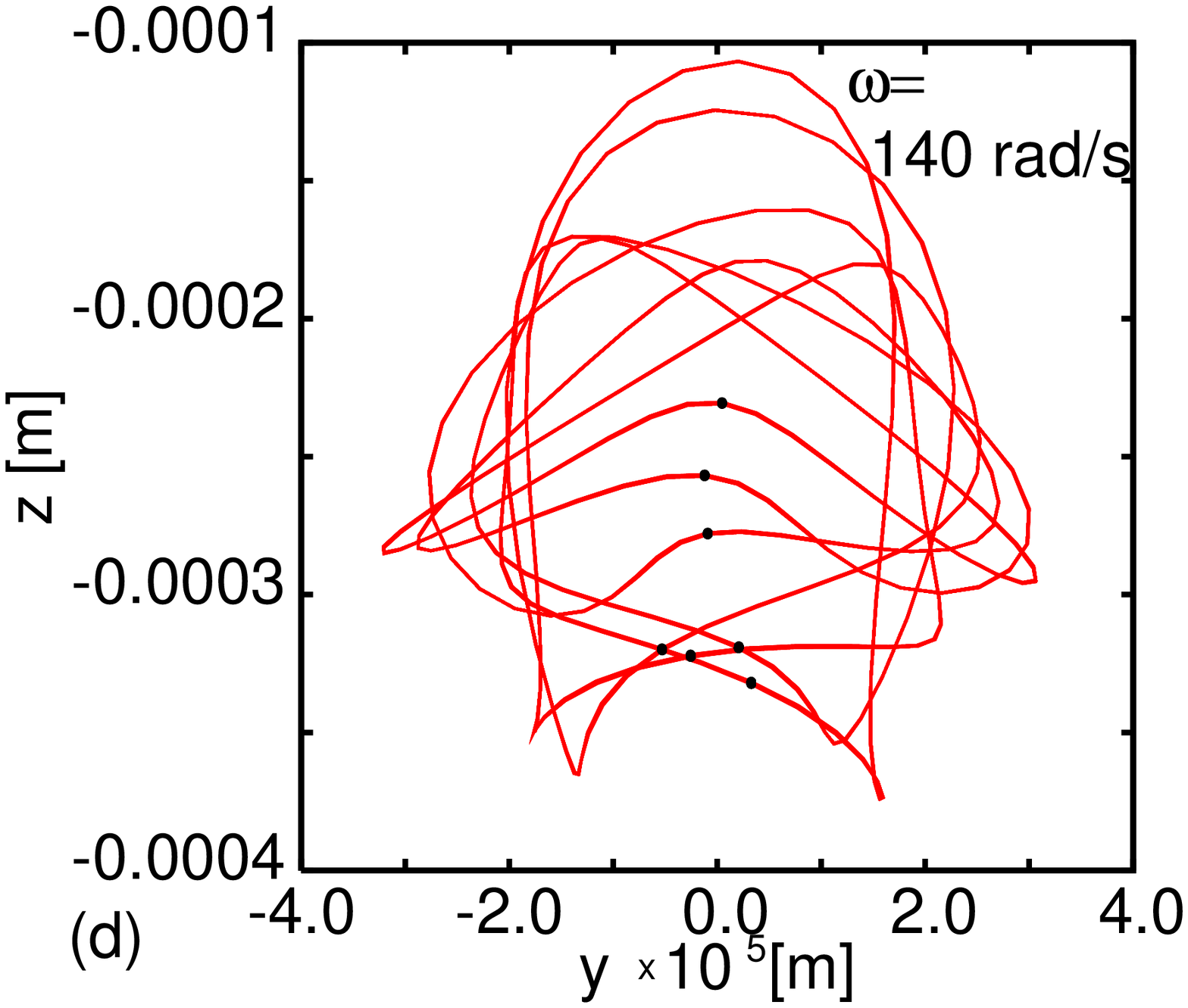}

\vspace{-0.3cm} 
\hspace{0.5cm}
\includegraphics[width=5.1cm,angle=-90]{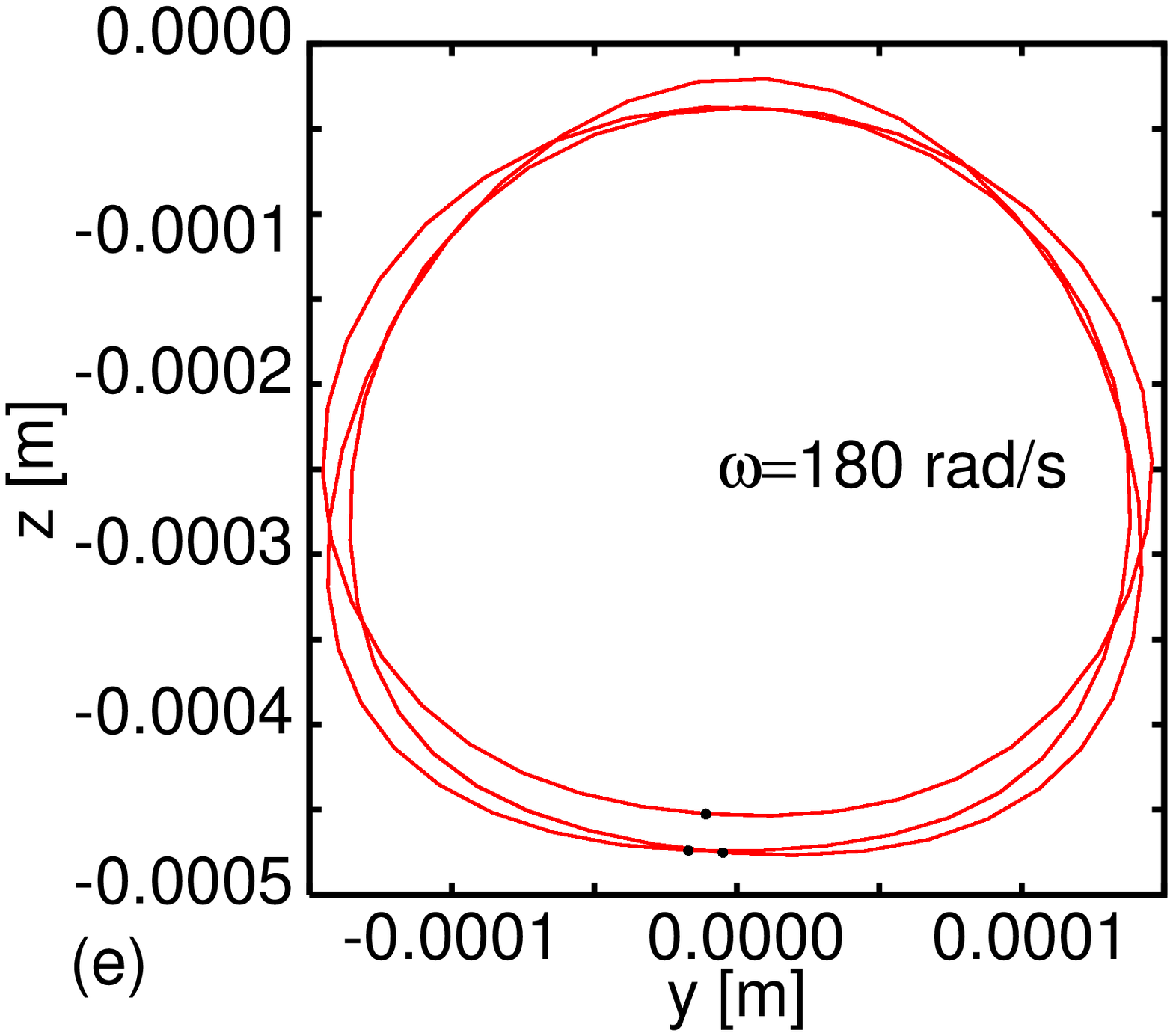} \hspace{-0.5cm}
\includegraphics[width=5.1cm,angle=-90]{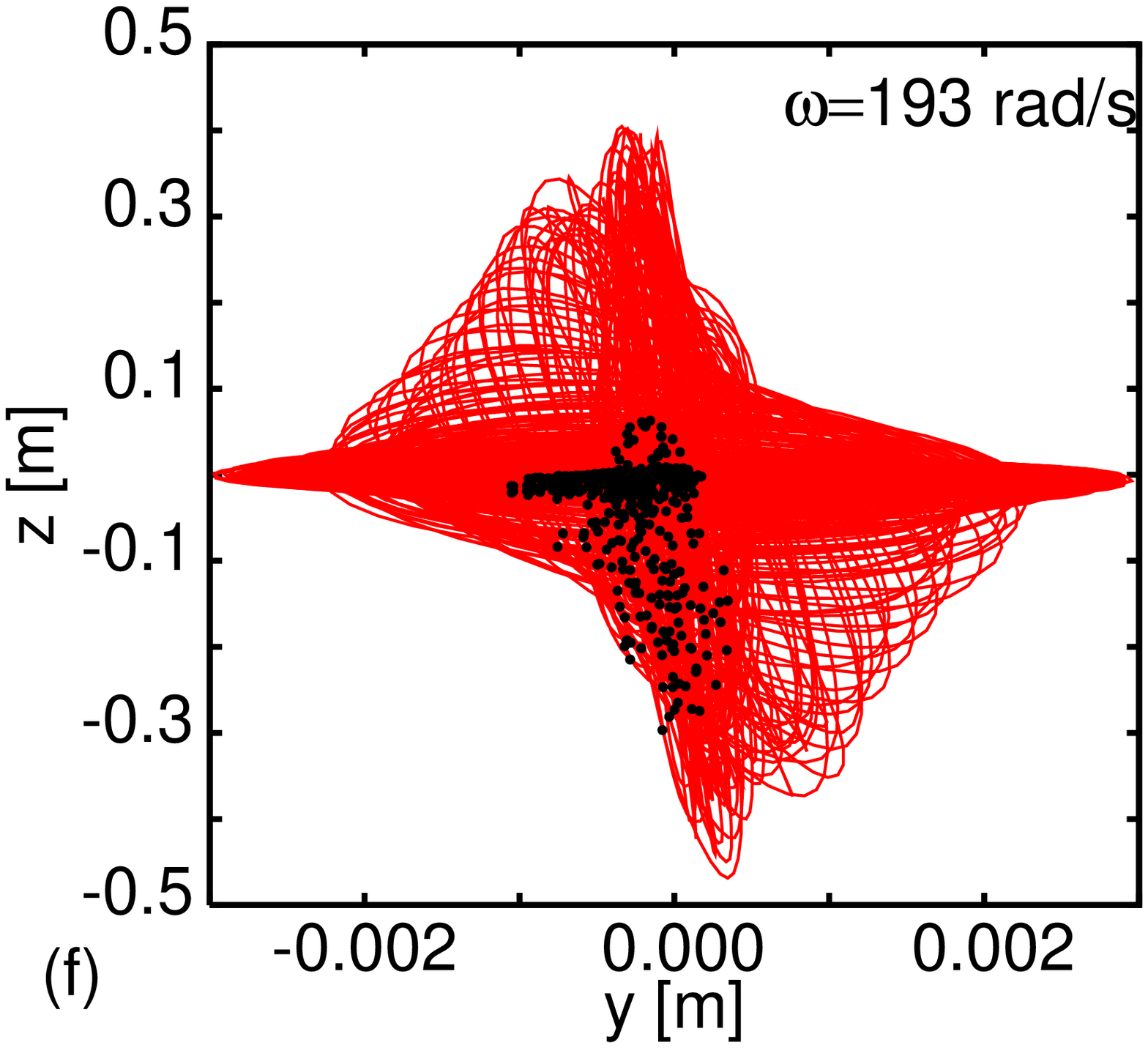}

\vspace{-0.3cm} 
\hspace{0.5cm}
\includegraphics[width=5.1cm,angle=-90]{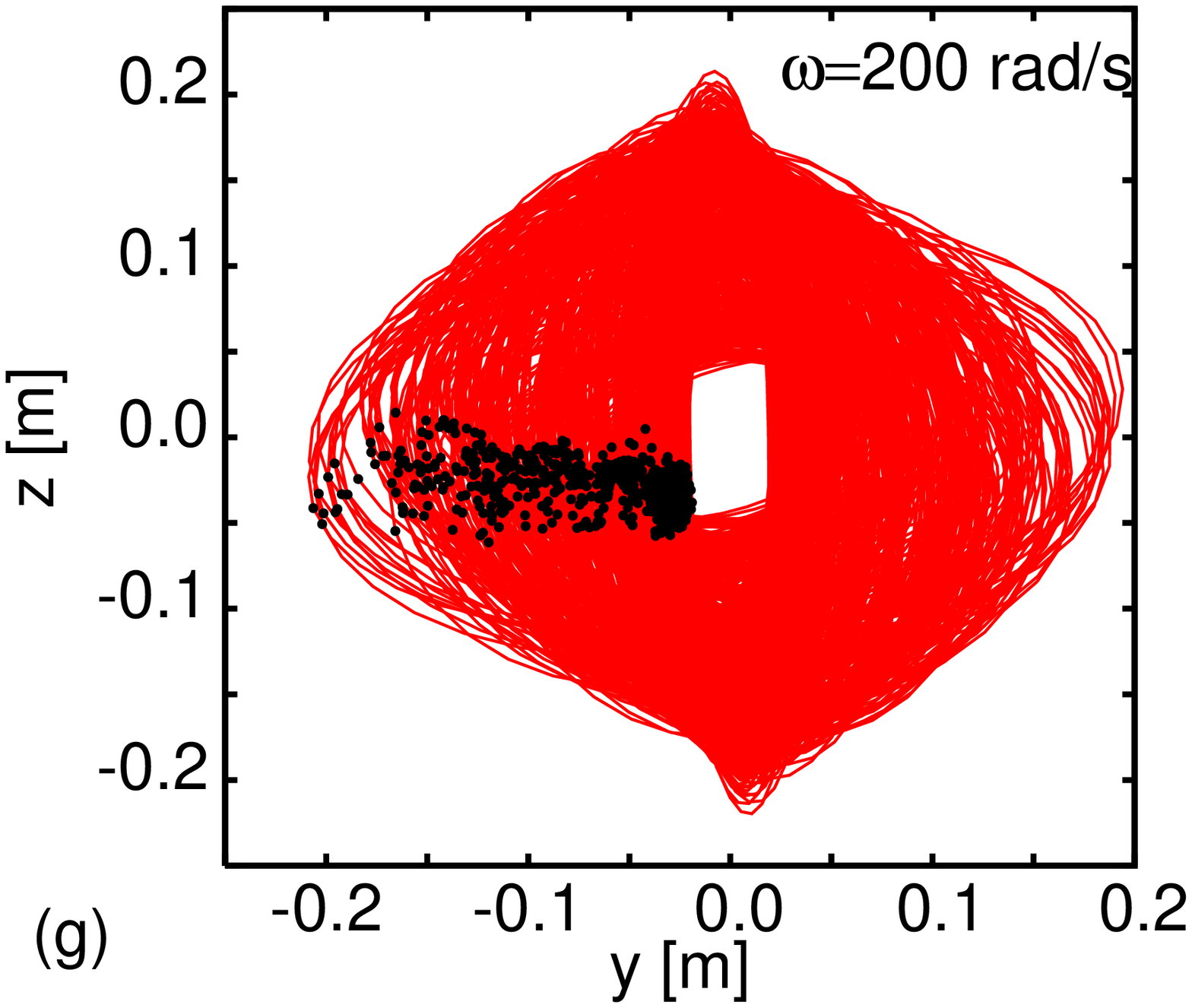} \hspace{-0.5cm}
\includegraphics[width=5.1cm,angle=-90]{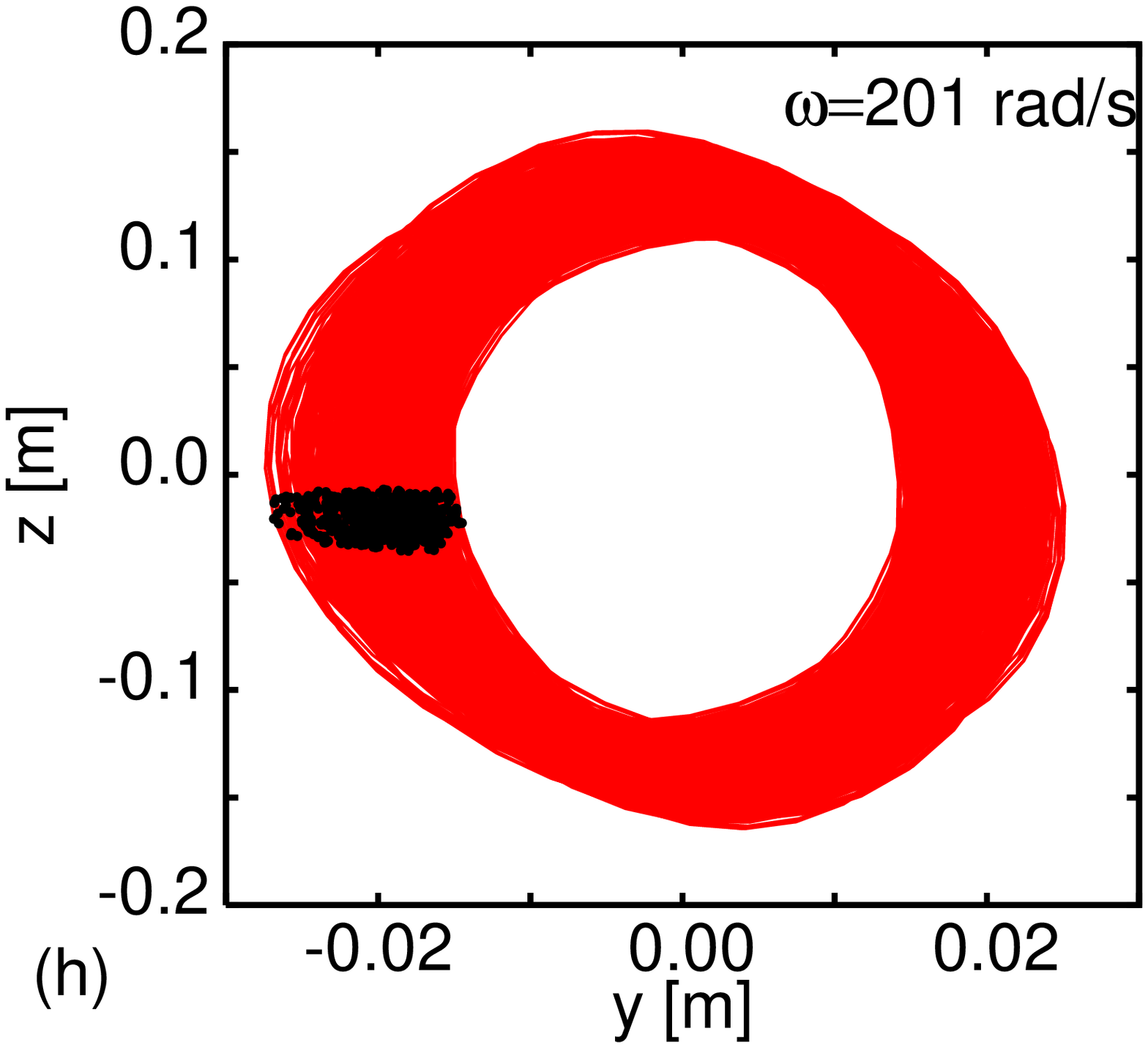}

\caption{ \label{fig4} Phase diagram projections into z-y plane for few values of speed velocities ($\omega=120$, 132, 133, 140,
180, 193, 200, and 201).
}
\end{figure}

\begin{figure}[htb]
\center{
 \includegraphics[width=9.0cm,angle=-90]{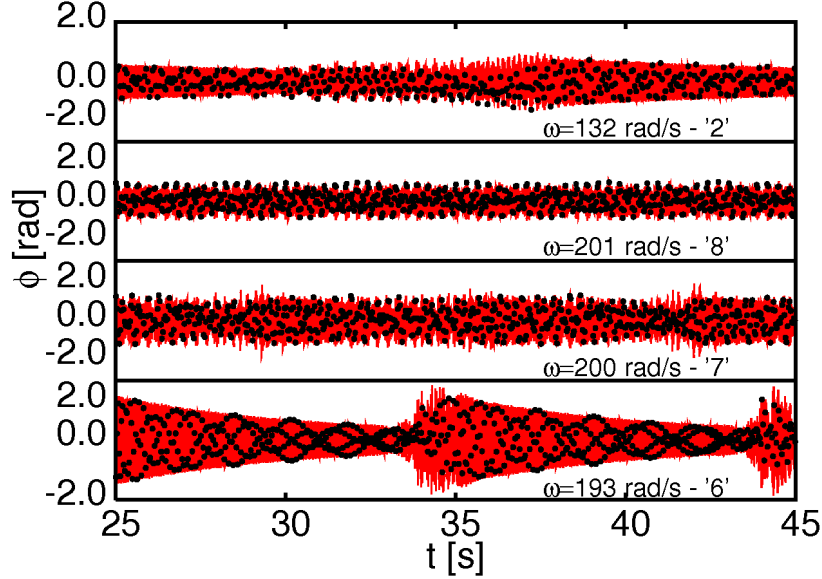}}

\caption{ \label{fig5} Time histories of torsional angular displacements $\phi$ for
intermittent oscillations
($\omega=132$, 193, 200, 201).
Black points shows stroboscopic
points.
Numbers '6', '7', '8', '2' indicate the corresponding values in the bifurcation diagram
 (Fig. 1).
}
\end{figure}

\begin{figure}[htb]

\vspace{-0.6cm} ~

\includegraphics[width=13.1cm,angle=0]{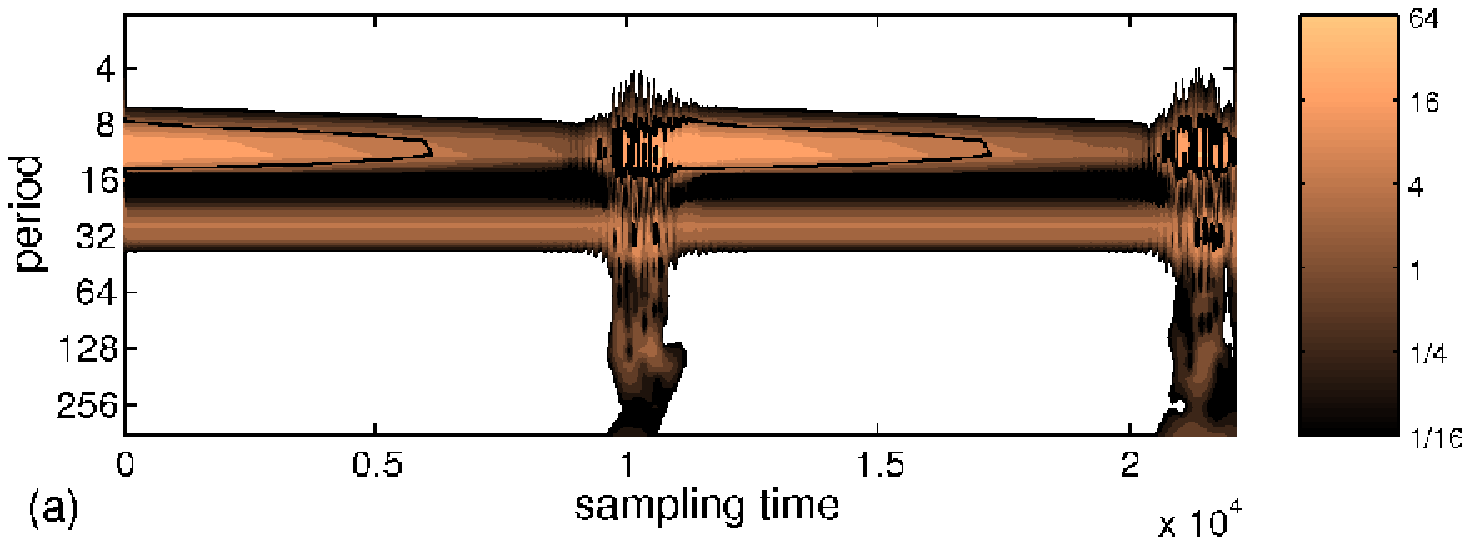}

\vspace{-0.6cm} ~

\includegraphics[width=13.1cm,angle=0]{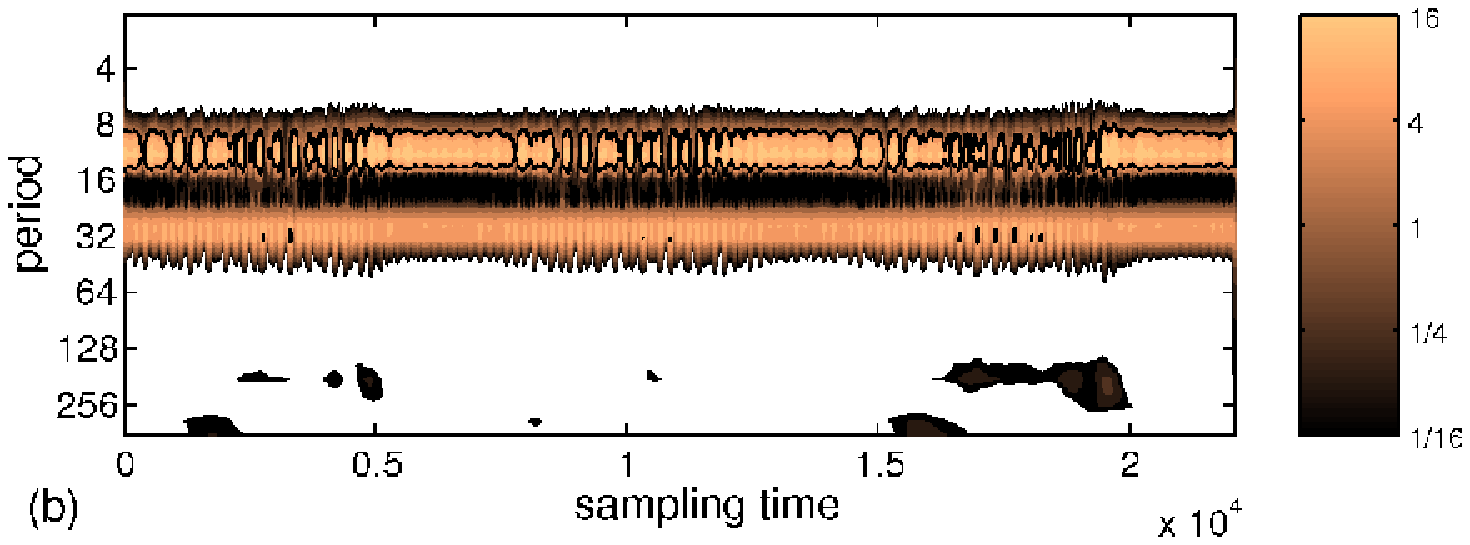}



\vspace{-0.6cm} ~

\includegraphics[width=13.1cm,angle=0]{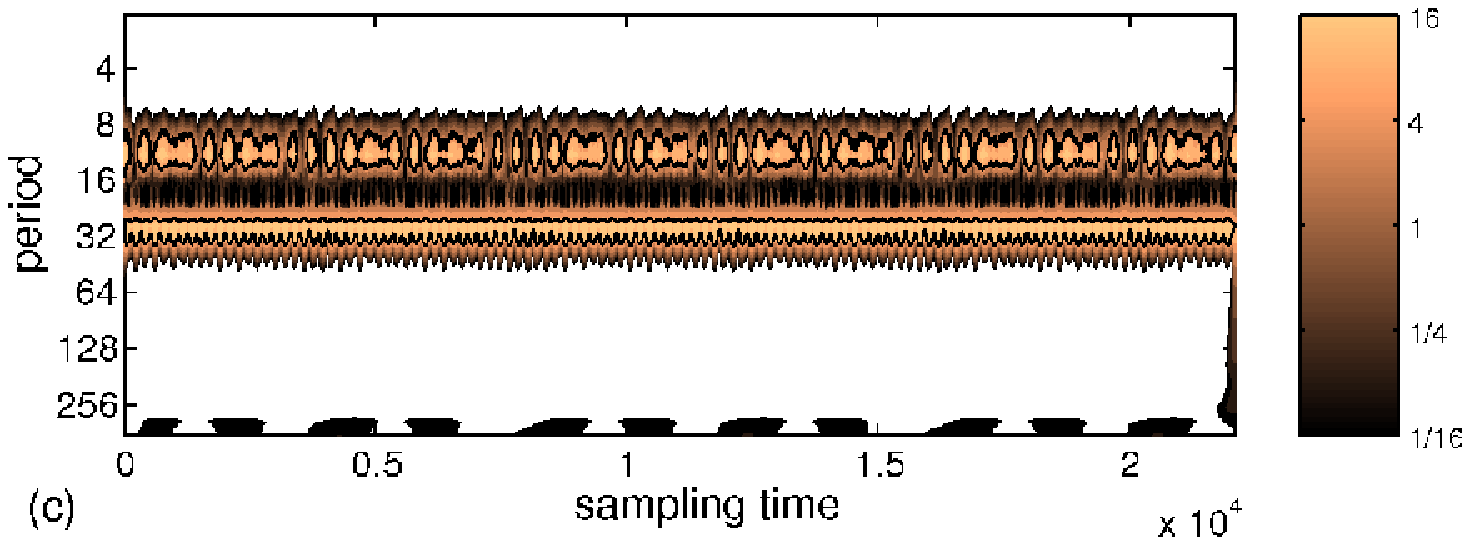}

\vspace{-0.6cm} ~

\includegraphics[width=13.1cm,angle=0]{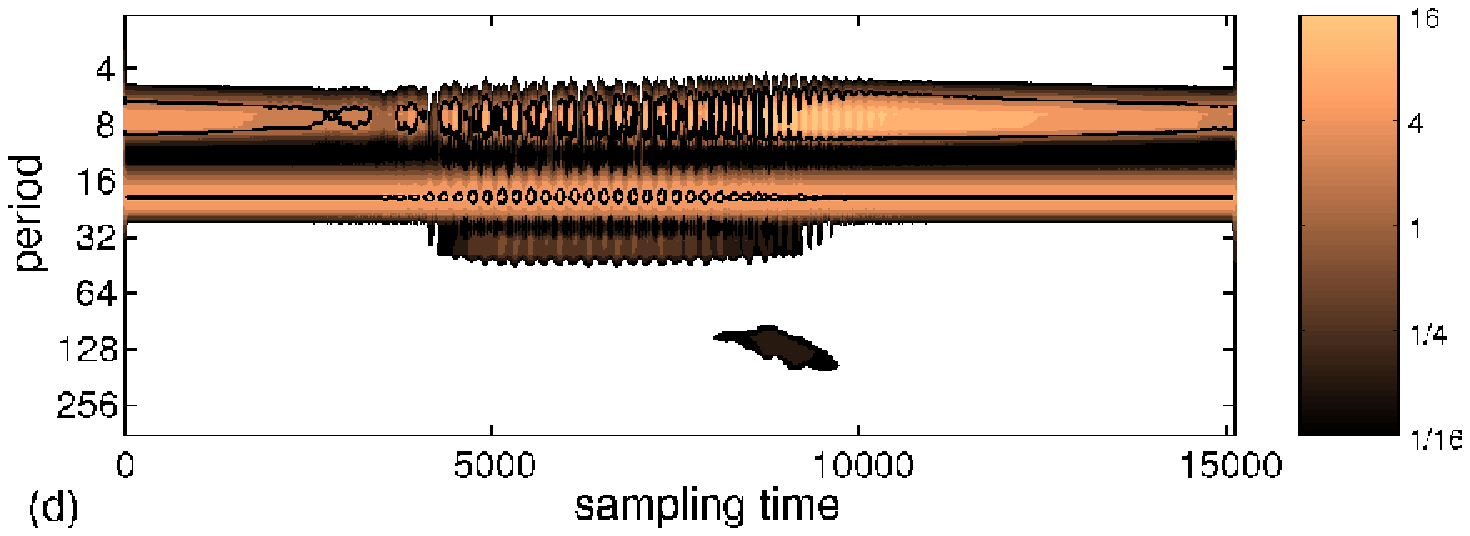}

\caption{ \label{fig6} Wavelet power spectra for torsional angular displacements $\phi$;
$\omega=193$, 200, 201, 132 for
(a), (b), (c), and (d); respectively. Note, the logarithmic colour 
scale on the right hand side. White regions in figures (a-d) indicate 
that power spectrum values exceed the scale.
}
\end{figure}

\begin{table}
\caption{
Parameters for a simple-supported Jeffcott rotor with a crack.
}
\vspace{0.5cm}
\begin{center}
{\small
\begin{tabular}{|c|c|}
\hline
Disk mass, $M$                         &   3 kg \\
Disk polar moment of inertia, $J_p$    &   0.01 kg m$^2$ \\
Eccentricity of the disk, $\epsilon$            &  2.2 $\times$ 10$^{-5}$ m \\
Torsional natural frequency, $\omega_t$ &          630 rad/s \\
Lateral natural frequency, $\omega_t$ &          210 rad/s \\
Shaft lateral stiffness, $k$  &  $k=\omega_l^2M$ \\
Shaft torsional  stiffness, $k_t$ & $k_t=\omega_t^2J_p$ \\
Damping ratio in lateral direction, $\xi_l$ ($C_l=2\xi_lM\omega_l$) &  0.008 \\
Damping ratio in torsional direction, $\xi_t$ ($C_t=2\xi_lJ_p\omega_t$)  & 0.0006 \\
External torque amplitude, $T_0$ & 800 Nm \\
External torque amplitude, $\omega_e$ & $\in [100,250]$ rad/s \\
Relative stiffness change caused by cracks  $\frac{\Delta k_{\xi}}{k}$ & 0.4 \\
($\Delta k_1=\frac{7}{6} \Delta k_{\xi}/k$, $\Delta k_2=\frac{5}{6} \frac{\Delta k_{\xi}}{k}$)
& ~~ \\
\hline
\end{tabular}}
\end{center}
\end{table}

The results for the range of rotational velocities 
$\omega \in [100,250]$ are presented in the 
bifurcation diagram  (Fig. 2). One can see that the resonances appear as 
they are expected
just below the lateral natural frequency and slightly above the 
half of lateral natural frequency, which
one would expect for a non-linear system of this kind. Spanning the rotor spin 
speed in the range of 145-149 rad/sec and
216-234 rad/sec, reveals the presence of a wide variety of the excited response 
characteristics. To explain the nature of vibrations in the regions of resonance we 
focused on the specific frequencies denoted by '1'--'9'
for the corresponding  values $\omega=$  120, 132, 133, 140, 180, 193, 200, 201, and 220 rad/s.
In Fig. 3a we show the time histories of displacement in the vertical direction  for 
nonresonat cases 
$\omega =$ 
120, 140,
220 rad/s. Black points are stroboscopic points plotted with the rotation frequency $\omega$.
Note that all three considered here cases indicate  periodic vibrations  with a small amplitude. 
  On the other hand in Fig. 3b we show the resonant time series  with  rotation 
velocities 
$\omega =$ 
132,
200, 201 
rad/s. In contrary to the previous cases the displacement $z$ can be large. Interestingly 
it increases after small laminar behaviour with a relatively small amplitude.  
Note that in both panels the numbers 
'1', '4',
'9', '2', '6', and '7' indicate the corresponding values $\omega$ in the 
bifurcation diagram
(Fig. 2).

To shed some more light into the lateral modes we plotted, in Figs. 4a-h
the 
phase diagram projections into z-y plane 
are plotted 
for few values of speed 
velocities( $\omega =$ 120, 132, 133, 140, 180, 193, 200, and 201).
Black points indicate the corresponding Poincar\'e maps.
In cases of nonresonant vibrations (Figs. 4a,d,e) of small amplitudes we observe 
with closed lines and a number of singular black points indicating periodic motion.
Interestingly in Fig. 4c  the Poincare map show a line instead of points indicating the 
quasiperiodic character of motion. The rest of plots (Fig. 4b,f,g) have non-periodic character. 
while Fig. 4h has features of moth quasi-periodic and non-periodic.  
Note that Fig. 4b,f,g, and h look 
like chaotic attractors, 
however to tell more about the 
non-periodic beats nature 
one should perform standard calculations of 
Lyapunov exponents or correlation dimension analysis.
    
Finally in Fig. 5 we  plot  the time histories of torsional angular displacements ( $\omega=$ 132, 
193, 200, 201). Black points 
shows stroboscopic points. 
Numbers
'6', '7', '8', '2' indicate the corresponding values in the bifurcation 
diagram (Fig. 1). These plots shows the relaxation nature of the torsional oscillations. 
This is consistent with results for intermittent lateral oscillations presented in Fig. 3.
Such an intermittent
 behaviour can
be
presented by temporal frequency diagram \cite{sen2007,sen2008} which can be given by the wavelet power spectra \cite{torrence1998}. In Fig. 6 one
clearly see the laminar (periodic)  interplay with turbulent (chaotic) regions.
  Note that for smaller rotation velocities we observe longer laminar  intervals.

\section{Discussion and Conclusions}
Using a simple model of the breathing crack we have showed that
the response to periodic torque applied at the ends of rotating shaft is complex.
 In our results we observe on one hand the multi-frequency synchronization, and 
on the other hand quasi-periodic motion and
nonlinear beats in the regions of resonances. Especially the beats are characteristic as far as the
nonlinear coupling of
lateral and torsional modes induced by the crack  is concerned.
The beats  are visible 
in both 
channels. They are also the signatures of interplay of external and parametric 
excitations, which 
arise from the crack existence. Interestingly the 
 beats causing the intermittency of the laminar phases
 introduce the non-periodic element of vibrations. The chaotic nature of such 
dynamic response of the rotor system 
exceeds the present report and will be discussed in detail in a future publication.    
 
\section*{Acknowledgment} GL would like to thank Dr. Michael Scheffler for helpful discussions.

\end{document}